\documentclass[11pt,a4paper]{article}
\usepackage{epsf,amsmath}
\usepackage{graphicx}

\begin{document}

\title{Inclusive and Associated $b$-jet Production \\at
the Tevatron in the Regge Limit of QCD}

\author{V.~A.~Saleev$^{a,b}$\footnote{{\bf e-mail}: saleev@ssu.samara.ru},
A.~V.~Shipilova$^{a}$\footnote{{\bf e-mail}:
alexshipilova@ssu.samara.ru}
\\
$^a$ \small{\em Samara State University} \\
\small{\em Ac. Pavlova, 1, Samara 443011, Russia}\\
$^b$ \small{\em Samara State Aerospace University} \\
\small{\em Moscow Highway, 34, Samara 443086, Russia }}
\date{}
 \maketitle

\begin{abstract}
\noindent We consider $b$-jet production in the
quasi-multi-Regge-kinematics approach based on the hypothesis of
the gluon and quark Reggeization in the $t$-channel exchanges at
the high energy. The data on various spectra of $b$-jet production
measured by the CDF and D0 Collaborations at the Tevatron Collider
are described well and with no free parameters.
\end{abstract}




\section{Introduction}
The study of $b$-jet production at high energies is of great
interest to test the perturbative quantum chromodynamics (QCD).
The presence of a heavy $b$ quark ($m_b\gg \Lambda_{\rm{ QCD}}$,
$\Lambda_{\rm QCD}$ --- asymptotic scale parameter of QCD)
guarantees a large momentum transfer that keeps the
strong-coupling constant small $\alpha_s(m_b)\leq0.1$. In
Refs.~\cite{CDF1,CDF2,D0} the modulus of the $b$-quark transverse
momentum $k_T\geq 32~\mbox{GeV}\gg m_b$, so that it is justified
to work in massless approximation and to assume beauty to be an
active flavor in the proton.

The total center-of-mass energy at the Tevatron,
$\sqrt{S}=1.96$~TeV in Run~II, sufficiently exceeds the scale
$\mu$ of the relevant hard processes $\sqrt{S}\gg \mu \gg
\Lambda_{\rm QCD}$. In this regime, the contributions to the
production cross section from subprocesses involving $t$-channel
exchanges of partons (gluons and quarks) may become dominant.
Then, the off-shell properties of the incoming partons can no
longer be neglected, and $t$-channel partons become Reggeized.
They can be described in quasi-multi-Regge kinematics (QMRK)
approach~\cite{QMRK}, based on an effective quantum field theory
implemented with the non-Abelian gauge-invariant action including
fields of Reggeized gluons ($R$)~\cite{Lipatov95} and quarks
($Q$)~\cite{LipatoVyazovsky}. In QMRK, the particles (multi-Regge)
or groups of particles (quasi-multi-Regge) produced in the
collision are strongly separated in rapidity. For the inclusive
$b$-jet production, this implies that a single $b$ quark is
produced in the central region of rapidity, while other particles,
including a $\bar b$ quark, are produced at large rapidities. In
the case of $b\bar b$ pair and $b\gamma$ associated production in
the central rapidity region, we also assume that there are no
other particles in this region, so that these particles are
considered as quasi-multi-Regge pairs. In the presented work we
continue our investigation and acknowledge the results obtained in
Refs.\cite{KniehlSaleevVasin1}--\cite{PRJpsi}.

\section{Inclusive $b$-jet production}
We examine inclusive single $b$-jet production in $p\bar p$
collisions in the fixed-flavor-number scheme with $n_f=5$ active
quark flavors. To leading order (LO) in the QMRK approach, there
is only one partonic subprocess $Q_b(q_1)+R(q_2)\to b(k)$, where
$Q_b$ is the Reggeized $b$ quark, and the four-momentum labels are
indicated in parentheses.
Using the effective vertex of the concerned transition given by
\cite{LipatoVyazovsky}, the squared amplitude of concerning
subprocess is found to be $\overline{|{\cal M}(Q_bR\to
b)|^2}=\frac{2}{3}\pi \alpha_s {\vec k}_T^2$~\cite{Saleev2008}.

At next-to-leading order (NLO) in the QMRK approach, the main
contribution to inclusive $b$-quark production arises from the
partonic subprocess $R(q_1)+R(q_2)\to b(k_1)+\bar b(k_2)$, where
the $b$ and $\bar b$ quarks are produced close in rapidity. The
contributions due to the other NLO processes, $RQ_b\to gb$,
$Q_q\bar Q_q\to b\bar b$, and $Q_q(\bar Q_q)Q_b\to q(\bar q)b$ are
suppressed because, in the small-$x$ region, the parton
distribution function (PDF) of the Reggeized gluon greatly exceeds
the relevant Reggeized quark PDFs. The squared amplitude of
subprocess~$RR\to b\bar b$ can be found in
Ref.~\cite{SaleevVasinBc}.

Exploiting the hypothesis of high-energy factorization, we write
the hadronic cross sections $d\sigma$ as convolutions of partonic
cross sections with unintegrated PDFs. We adopt the
Kimber-Martin-Ryskin prescription \cite{KMR} for unintegrated
gluon and quark PDFs, using as input the
Martin-Roberts-Stirling-Thorne collinear PDFs of the proton
\cite{MRST}.

In Fig.~\ref{fig:1}, the preliminary CDF data \cite{CDF1} are
compared with our predictions obtained in the QMRK approach. The
contributions due to LO and NLO subprocesses are shown separately.
Since the lower bound of the $k_{2T}$ integration is zero, we
allow for the $b$-quark mass to be finite, $m_b=4.75$~GeV. The
renormalization and factorization scales are identified and chosen
to be $\mu=\xi k_T$. Here and later $\xi$ is varied between 1/2
and 2 about its default value 1 to estimate the theoretical
uncertainty, and the resulting errors are indicated in figures as
shaded bands. We observe that the contribution due to LO
subprocess greatly exceeds the one due to NLO subprocess, by about
one order of magnitude, and practically exhausts the full result.
It nicely agrees with the CDF data throughout the entire $k_T$
range.

\section{$b\bar b$-pair production}
\begin{figure}[!thb]
\vspace*{16cm}
\begin{center}
\includegraphics{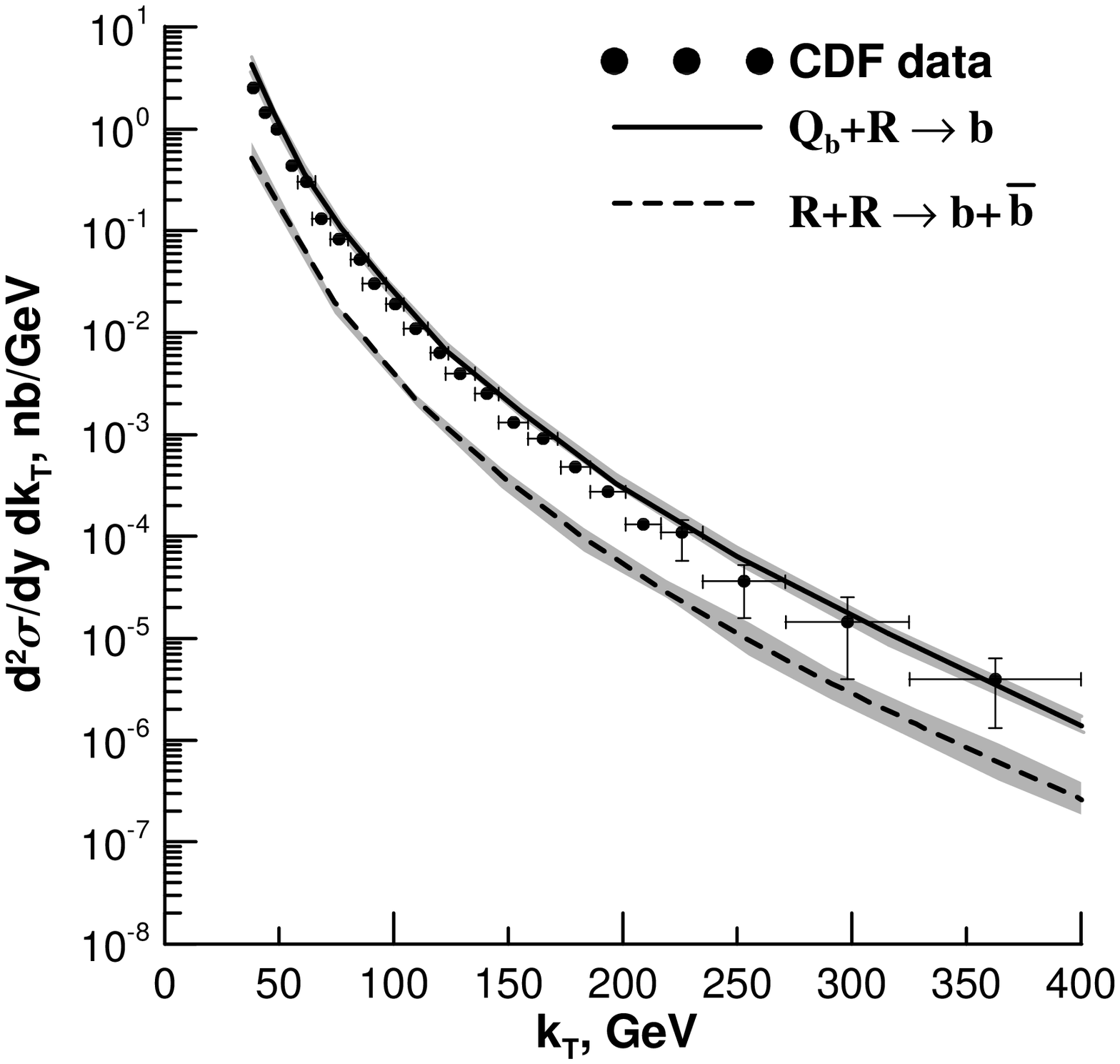}
\includegraphics{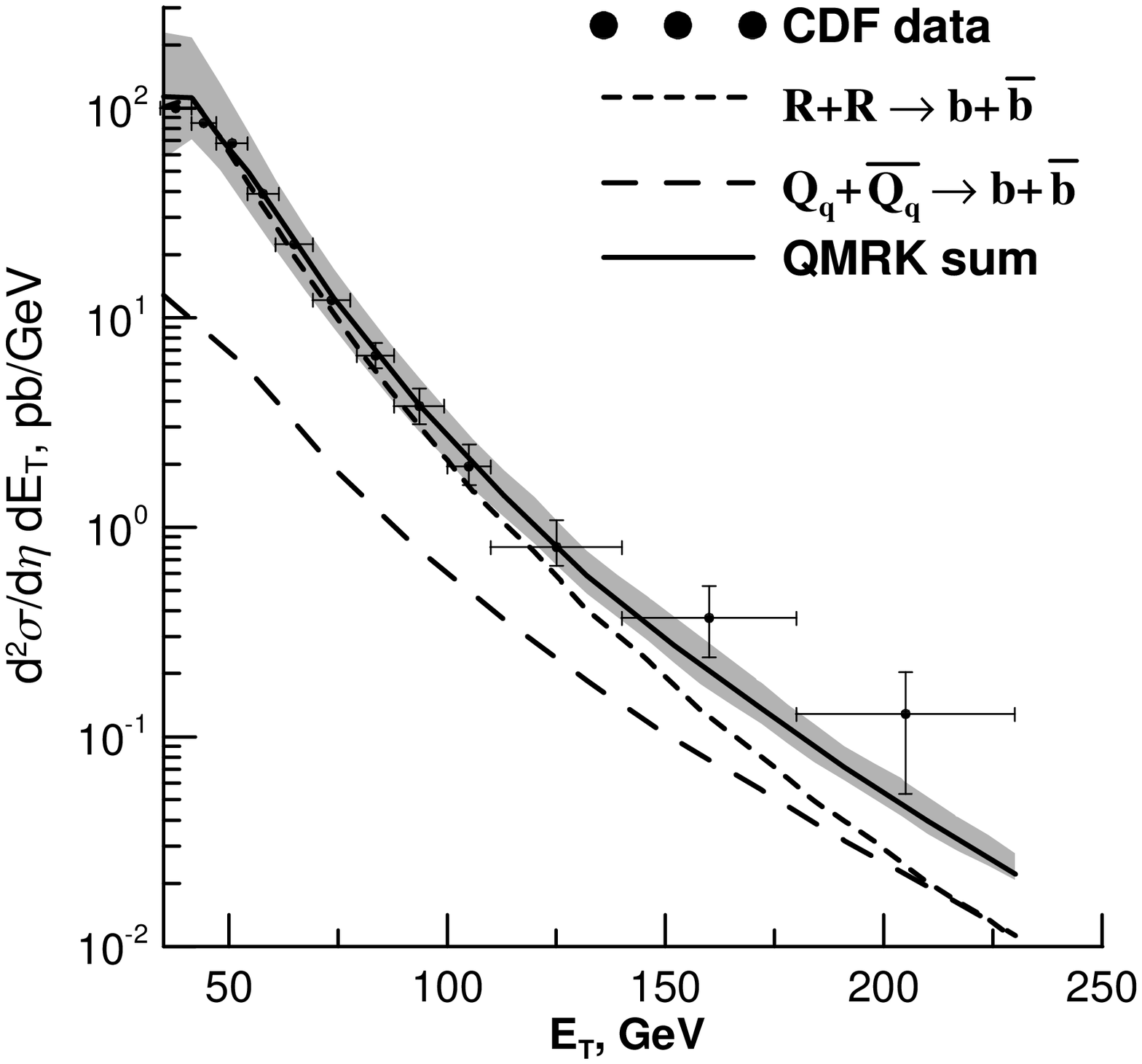}
\includegraphics{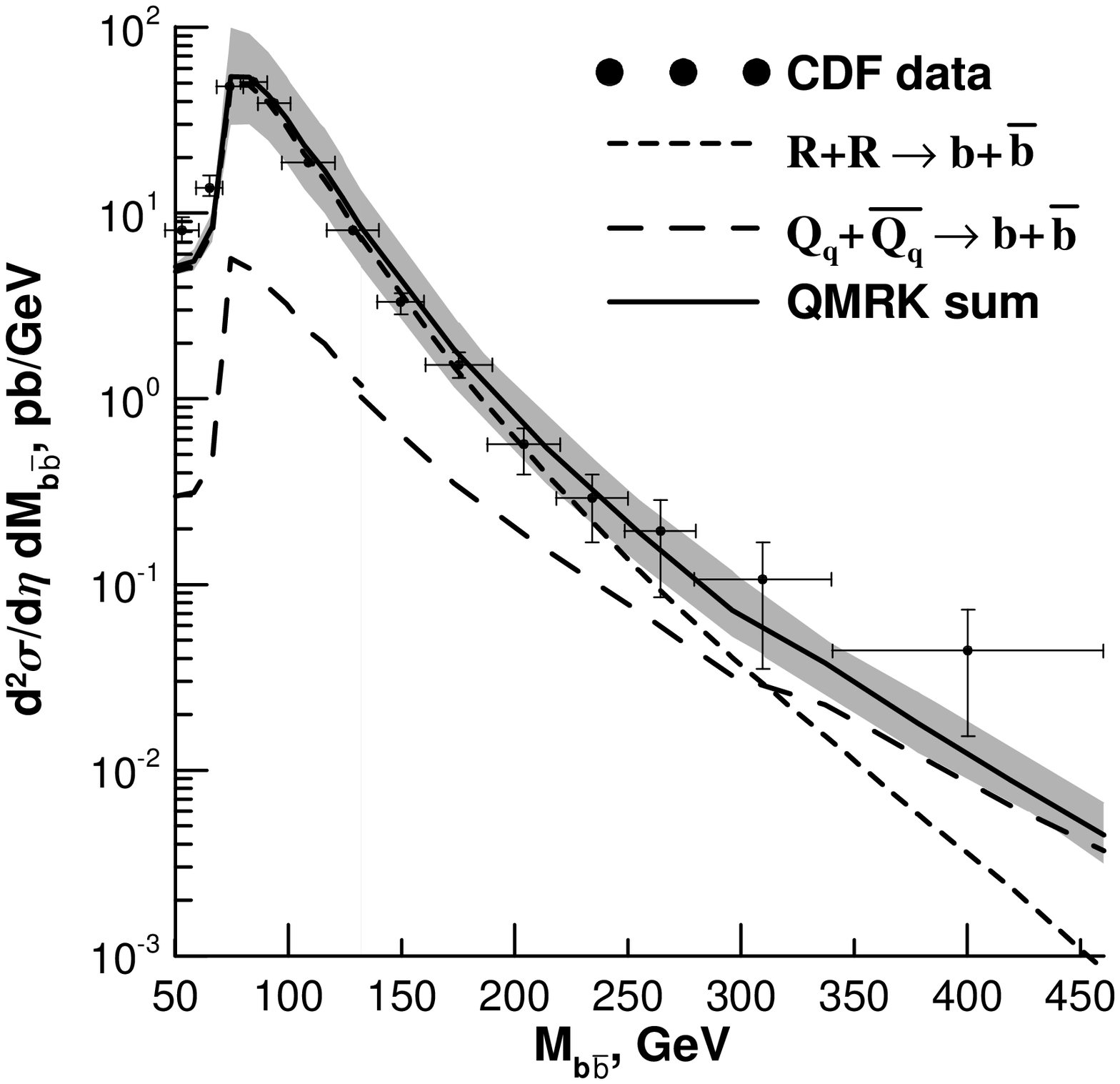}
\includegraphics{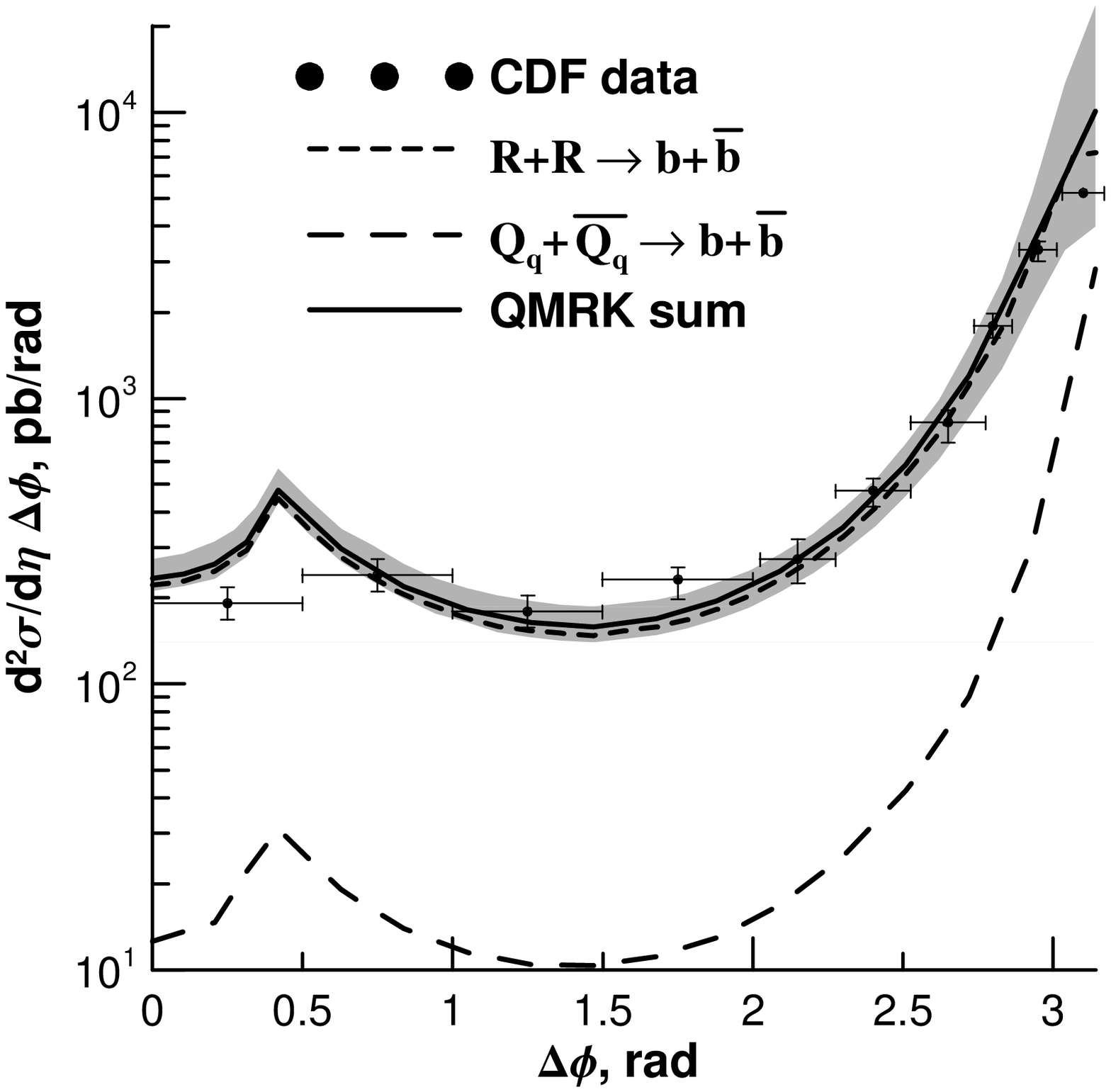}
\caption[*]{\label{fig:1} The distribution in (left, top)
transverse momentum of inclusive single $b$-jet
hadroproduction~\cite{CDF1}; the ones in (right, top) leading-jet
transverse energy, (left, bottom) dijet invariant mass, and
(right, bottom) azimuthal separation angle of inclusive $b\bar
b$-dijet hadroproduction~\cite{CDF2} are compared with the QMRK
predictions.}
\end{center}
\end{figure}

The data measured by the CDF Collaboration on inclusive $b\bar
b$-dijet production cross section \cite{CDF2} come as
distributions in transverse energy $E_{1T}$ of the leading-jet
(jet with the maximal transverse energy), the dijet invariant mass
$M_{b\bar b}$, and the azimuthal separation angle $\Delta\phi$.
They are compared with our QMRK predictions including the
contributions from subprocesses $RR\to b\bar b$ and $Q_q\bar
Q_q\to b\bar b$, where $q=u,d,s,c$, in Fig.~\ref{fig:1}. The
common scale is set to be $\mu=\xi k_{1T}$.
In Fig.~\ref{fig:1} these two contributions are shown separately
along with their superpositions. We observe that the total QMRK
predictions nicely describe all the three measured cross section
distributions. The contributions due to subprocess~$RR\to b\bar b$
dominate for $E_{1T}\leq200$~GeV and $M_{b\bar b}\leq300$~GeV and
over the whole $\Delta\phi$ range considered. The peak near
$\Delta\phi=0.4$ arises from the isolation cone condition
\cite{CDF2}.

\section{Associated $b\gamma$ production}

The D0 Collaboration presented data on photon associated heavy
quark ($b,c$) production
~\cite{D0}, which can be realized via two mechanisms: the direct
photon production and the fragmentation of the final partons into
photons. In the former case, to leading order (LO) in the QMRK
approach, there is only one partonic subprocess $Q_bR \to b
\gamma$. The second mechanism of prompt photon production is the
fragmentation of the produced partons into photons, which can be
described by the perturbative fragmentation
functions~\cite{FFrag}. In this case  we need to take into account
partonic subprocesses with at least one heavy quark in the final
state: $RR\to b\bar b$, $Q_q \bar Q_q \to b \bar b$, $Q_b\bar Q_b
\to b \bar b$, $Q_b Q_q \to b q$, and $Q_b Q_b \to b b$. The
effective vertices of the concerned transitions are given by
\cite{LipatoVyazovsky} and the squared amplitudes can be found in
Refs.~\cite{KSS2010,by}.

\begin{figure}[!thb]
\vspace*{16cm}
\begin{center}
\includegraphics{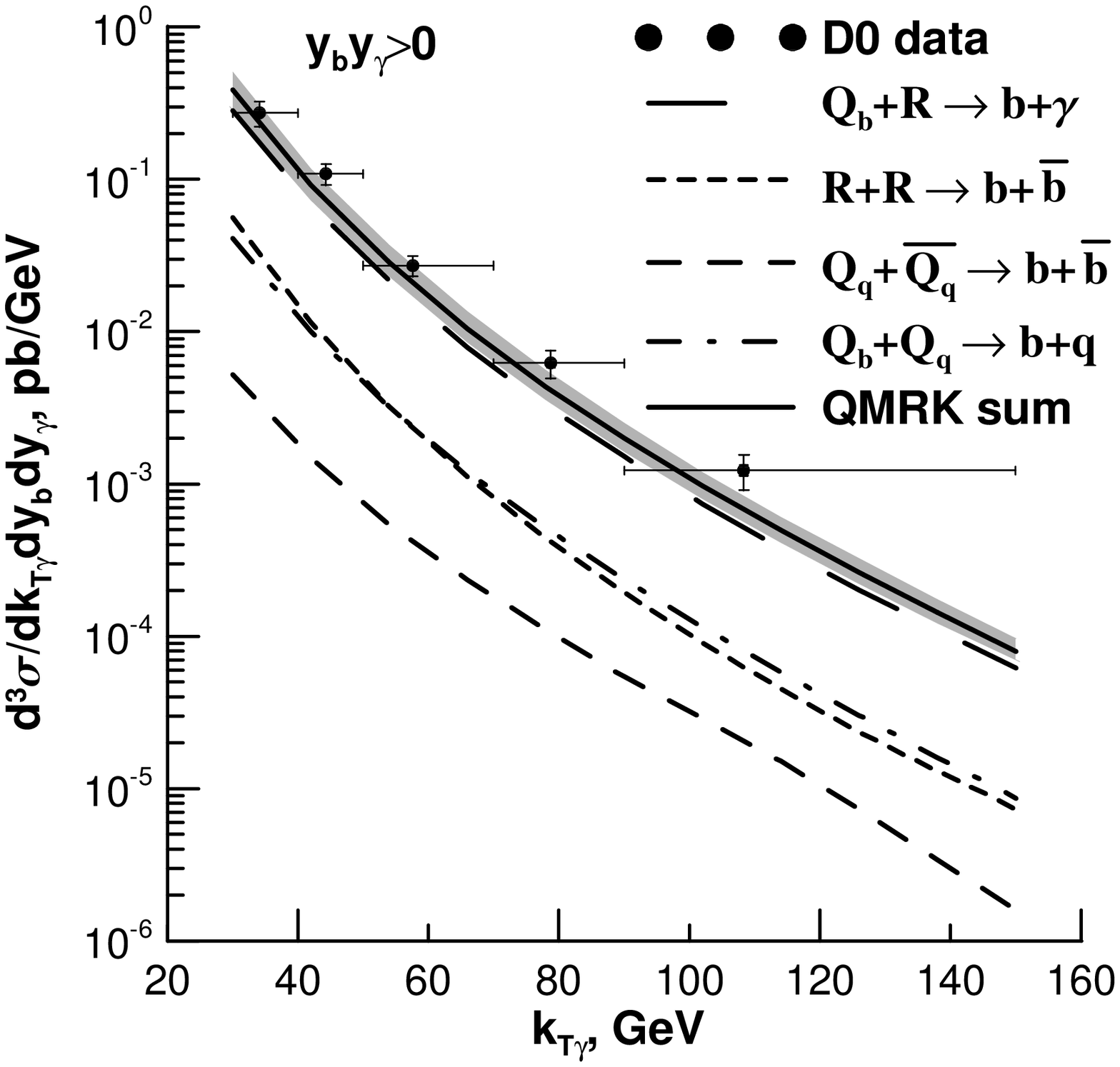} \includegraphics{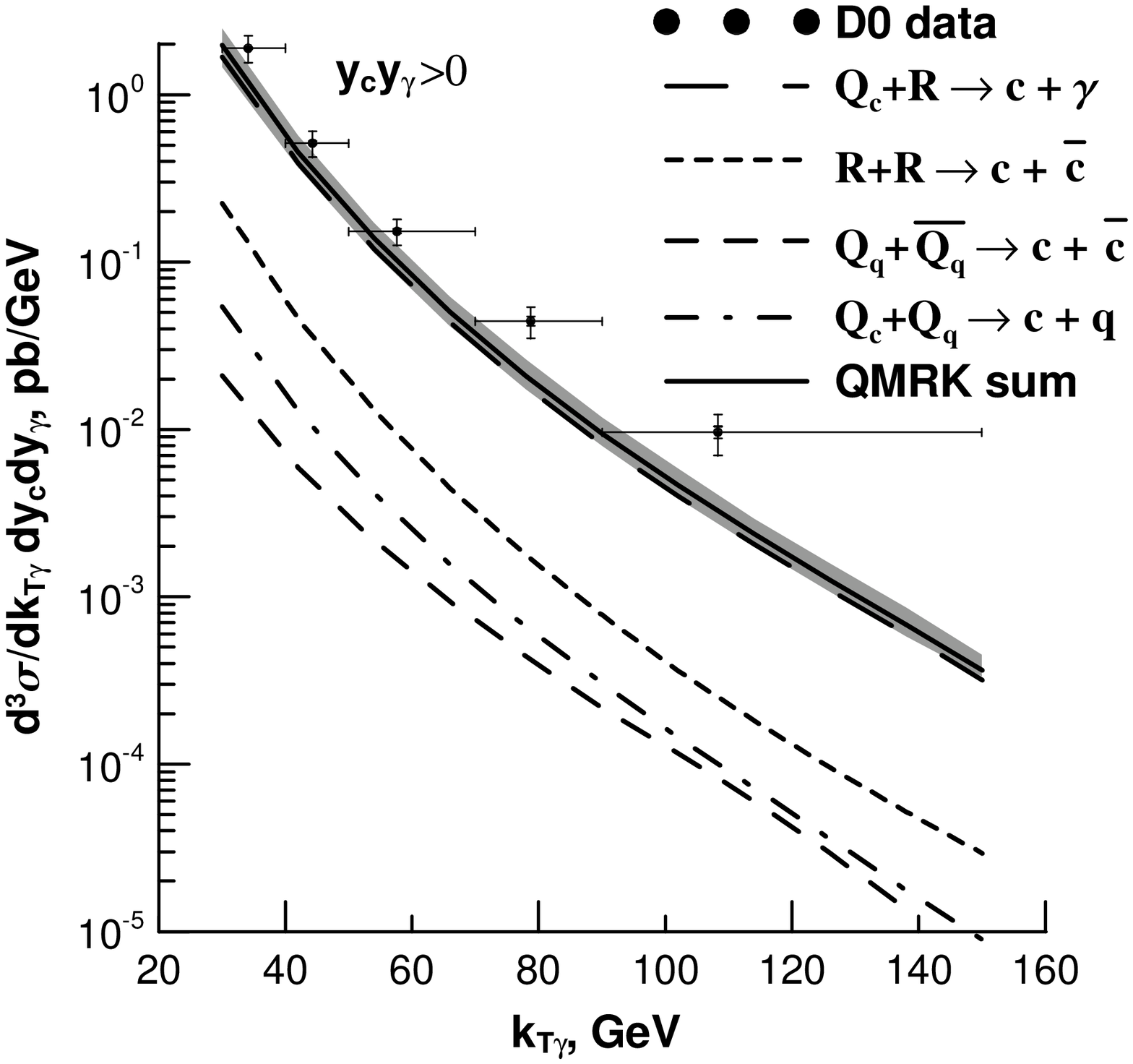}
\includegraphics{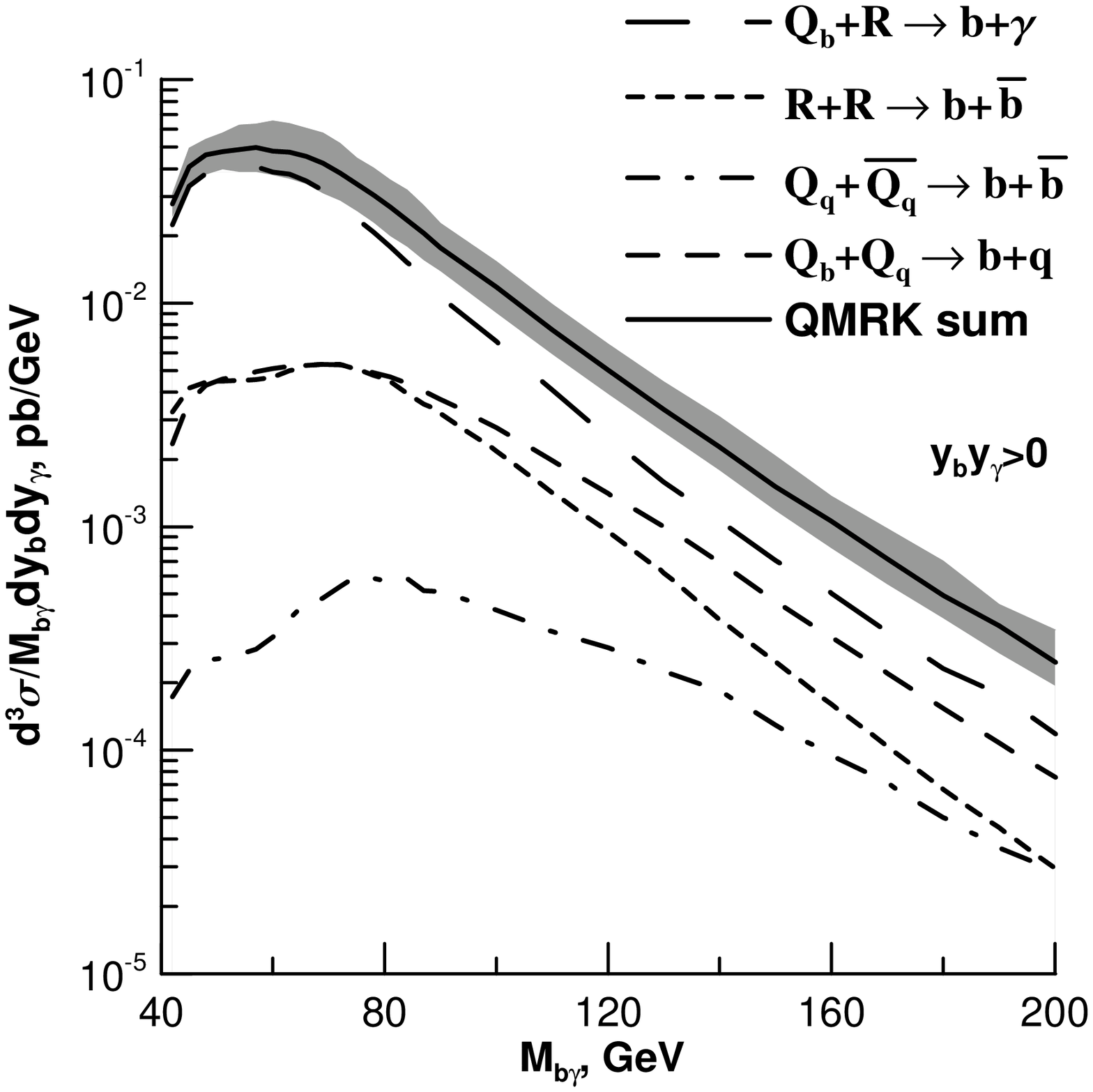} \includegraphics{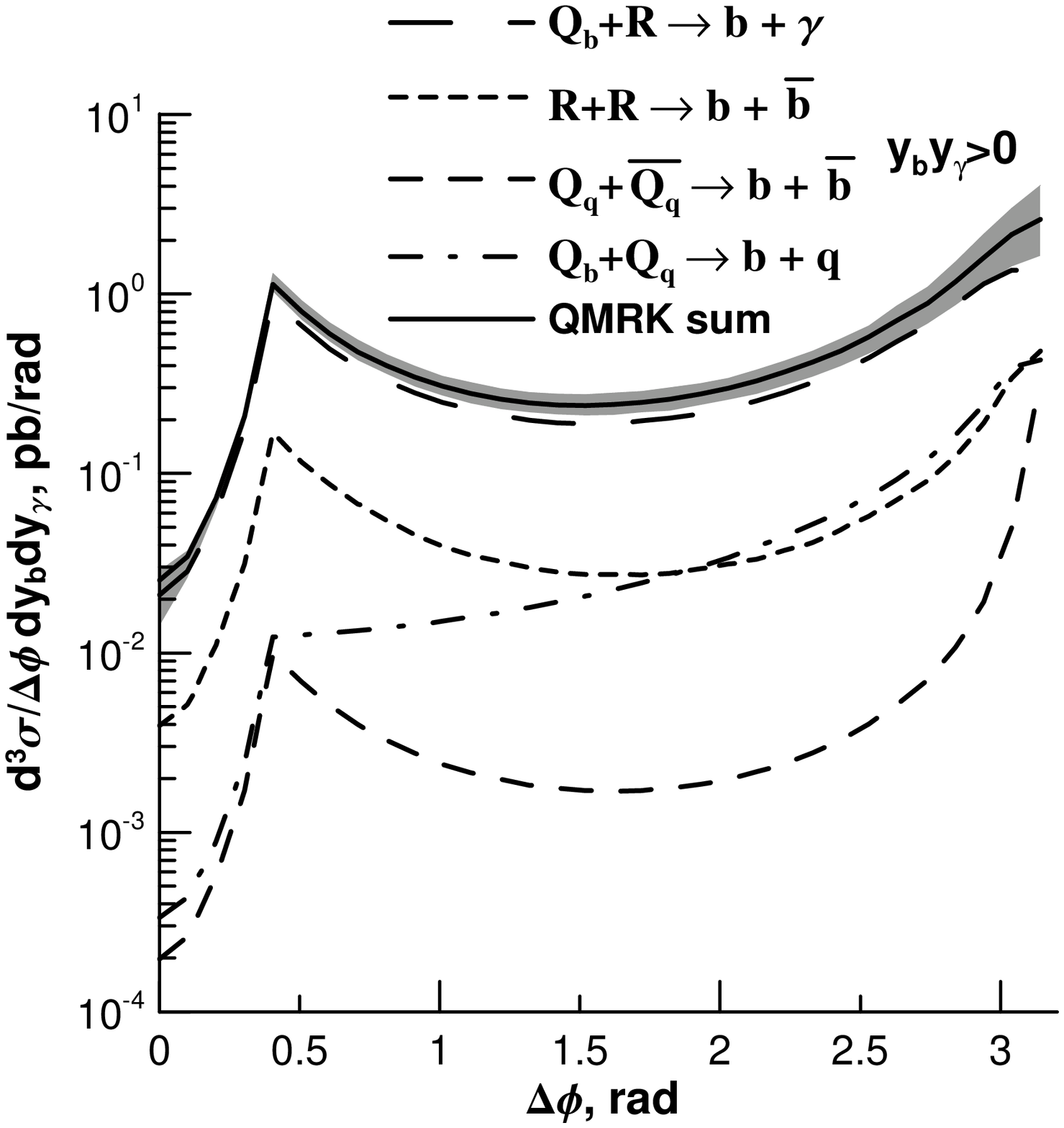}
\caption[*]{\label{fig:2} The QMRK predictions in comparison with
distributions in transverse momentum of associated $b\gamma$
(left, top) and $c\gamma$ (right, top) hadroproduction at
$y_{b(c)y_\gamma>0}~\cite{D0}$; the ones for distributions in
(left, bottom) $b\gamma$ invariant mass, and (right, bottom)
$b\gamma$ azimuthal separation angle.}
\end{center}
\label{fig2}
\end{figure}

In Fig.~\ref{fig:2}, top, the D0 data for $b$ and $c$ quarks are
compared with our QMRK predictions with $\mu=\xi k_{\gamma T}$.
We observe that the contribution due to direct photon production
greatly exceeds the one due to fragmentation photon production, by
about of one order of magnitude at large $k_{\gamma T}>40$~GeV and
by a factor 5 at the $k_{\gamma T}\simeq 30$~GeV. The direct
photon contribution practically exhausts the full result. It
nicely agrees with the D0 data throughout the entire $k_{\gamma
T}$ range for $b\gamma$ production and in the region of $k_{\gamma
T}\leq60$ GeV for $c\gamma$ production, while in the remaining
region for the latter our prediction underestimates the data by
factor about 2. We don't observe any difference between
description of the two kinematical ranges, distinguished by the
experiment $|y_{b,c}y_\gamma|>0$ and $|y_{b,c}y_\gamma|<0$
\cite{D0}.

We also present our predictions for the $b(c)\gamma$ invariant
mass and the azimuthal separation angle $\Delta\phi$ distributions
in Fig.~\ref{fig:2}, bottom.

\section*{Acknowledgements} The work of V.A.S. and A.V.S. was supported in
part by the Ministry of Science and Education of the Russian
Federation under Contract No.~P1338. The work of A.V.S. was also
supported in part by the International Center of Fundamental
Physics in Moscow and the Dynastiya Foundation.

\end{document}